\documentclass[pra,aps,showpacs,twocolumn]{revtex4}
\usepackage{graphicx}
\usepackage{dcolumn}
\usepackage{amsmath}
\begin{document}
\title{Double jumps and transition rates for two dipole-interacting atoms}
 \author{S\"oren U. Addicks$^1$}
\author{ Almut Beige$^2$}
\author{ Mohammed Dakna$^1$}  
\author{Gerhard C. Hegerfeldt$^1$}
\affiliation{$^1$Institut f\"ur Theoretische Physik, 
Universit\"at G\"ottingen,\\
Bunsenstr. 9, 73073 G\"ottingen, Germany\\
$^2$Optics Section, Blackett Laboratory, Imperial College London,\\
London SW7 2BZ, England}
\begin{abstract}
Cooperative effects in the fluorescence of two dipole-interacting
atoms, with macroscopic quantum jumps (light and dark periods), are 
investigated. The transition rates
between different intensity periods are calculated in closed form and
are used to determine the rates of double jumps between periods of
double intensity and dark periods, the mean duration of the 
three intensity periods and the mean rate of their occurrence. We
predict, to our knowledge for the first time, 
cooperative effects for double jumps, for atomic
distances from  one and to ten wave lengths of the strong 
transition. The double jump rate, as a function
of the atomic distance, can show oscillations of up to  $30\%$ at
distances of about a wave length, and oscillations are 
still noticeable at a distance of ten wave lengths.  The cooperative
effects of the quantities and their
characteristic behavior turn out to be strongly dependent on the
laser detuning.   
\end{abstract}
\maketitle
\section{Introduction}\label{intro}

The dipole-dipole interaction between two atoms can be understood
through the exchange of virtual photons and depends on the
transition dipole moment of the levels involved. It can be
characterized by  complex coupling constants, or by their real and
imaginary parts, where the former 
affect decay constants and the latter  lead to level
shifts \cite{aga}. Cooperative effects in the radiative behavior of atoms
which may arise from their mutual dipole-dipole interaction 
have attracted considerable interest in the
literature \cite{aga}-\cite{Ho}. 
 Two of the present
authors \cite{BeHe4} have  investigated in detail the transition from
anti-bunching to bunching with decreasing atomic distance for two
dipole-dipole interacting two-level atoms.

The striking phenomenon of macroscopic quantum jumps (electron
shelving or macroscopic dark and light periods) can occur for a
multi-level system where the 
electron  is essentially shelved for 
seconds or even minutes in a metastable state without photon emissions
\cite{SauterL}-\cite{HePle1}. For two such systems the
fluorescence behavior would, without cooperative effects, be just the
sum of the separate photon emissions, with dark periods of both atoms,
light periods of a single atom and of two atoms. In Ref. \cite{Sauter} the
fluorescence intensity of three such ions in a Paul trap was
measured and a large fraction of double or triple jumps  was reported,
i.e.   jumps by two or three intensity steps within the short
resolution time. This fraction was orders of 
magnitudes larger 
than that expected for independent ions. A quantitative explanation of
such a 
large cooperative effect for distances of the order of ten wave lengths 
of the strong transition has been found to be difficult
\cite{hendriks,Java,Aga88,Lawande89,Chun}. Other experiments at
larger distances and with different ions showed no cooperative
effects  \cite{Itano88,Thom}.

Quite recently, two of the present authors \cite{BeHe5} investigated
 for two such systems cooperative effects  in the mean
duration, $T_0$, $T_1$, and $T_2$, of the  dark,
single-intensity, and double-intensity periods, respectively. This was
done by simulations for two atoms in a $V$~configuration. The mean
duration of the single- and double-intensity periods 
depended sensitively on the dipole-dipole interaction and thus on the atomic
distance $r$. They exhibited noticeable oscillations which decreased in
amplitude when $r$ increased. These oscillations seemed to continue up
to a distance of well over five wave lengths of the strong transition
and they were {\em opposite}\/ in phase with those of Re$\,C_3(r)$,
where  $\,C_3$ is the complex dipole-dipole coupling constant
associated with the strong transitions.

In this paper we present an analytic approach to study cooperative
effects for atoms with macroscopic quantum jumps. This is explained
for two atoms, but is easily generalized. The approach is based on an
explicit calculation of transition rates between the various intensity
periods. From the transition rates all interesting statistical
quantities can be determined, such as double jump rates and mean
duration of different periods.

 We predict, to our knowledge for the
first time, cooperative effects in the double jumps
of two dipole-dipole interacting atoms. These results are  for
atoms in the $V$ configuration (see Fig.~\ref{V-system}) and are
verified  by simulations. As a function of the 
atomic distance, the  double jump
rates show marked oscillations, with a maximal difference of
up to 30\%,  decreasing as $1/r$. Most surprising is a change in
the oscillatory behavior of the double jump rate from {\em in phase} with
Re$\,C_3(r)$ to {\em opposite} in phase  when the detuning of the laser
driving the weak atomic 
transition is increased. For the mean durations $T_1$ and $T_2$ there
can be a change in behavior from opposite in phase  to in phase with
Re$\,C_3(r)$.  Moreover, for a particular value of the detuning,
which depends on  the other parameters, the double jump rate becomes
{\em constant} in  $r$ and the cooperative effects disappear. This is 
true  also for the mean period
durations and for their mean occurrences, with  different values of the
detuning, though. Typically, for nonzero detuning the oscillation
amplitudes do not exceed those found for zero detuning. 

The experiments of Ref. \cite{Sauter} exhibited extremely large
cooperative effects, in fact up to three orders of magnitude. Since
this was for a different atomic level configuration and for {\em three}
\/ions in a trap our results do not apply directly. In principle,
however, our analytic approach can be carried
over to the experimental situation of Ref. \cite{Sauter}, although the
calculations become algebraically more involved and have not been
carried out so far. 

The plan of the paper is as follows.  In Section~\ref{pij} 
 the fluorescence with its three different 
intensity periods is treated as a three-step telegraph process and the
Bloch equations are used to derive the
transition rates between the periods. In 
Sections~\ref{DJTheory} and \ref{duration} expressions for the double
jump rate and the mean duration of the three types of  intensity 
periods are obtained  
by means of these transition rates. The results are compared with 
simulations in which photon intensities are obtained by averaging
photon numbers over a small time window. It turns out that this
data-smoothing procedure can 
affect the results, and we show how this  can  be corrected for
quantitatively. A similar effect  can 
also occur when photon detectors measure the intensity of light by
averaging over a small time window. 
 In the last section the results are discussed.
It is suggested that the mean rate of double-intensity periods is
an experimentally  more easily accessible candidate for exhibiting
cooperative effects arising from the dipole-dipole interaction. 

\section{Transition rates}\label{pij}
\subsection{Prerequisites}
\label{prereq}
\begin{figure}[bth]
\includegraphics[width=3in]{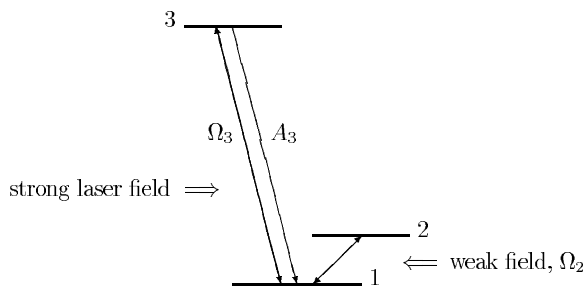}
\caption {V system with metastable level 2 and Einstein
coefficient $A_3$ for level 3. $\Omega_2$ and $\Omega_3$ are the Rabi
frequencies of the two lasers driving the weak 1-2 transition and the
strong 1-3 transition, respectively.
\label{V-system}}
\end{figure}

\begin{figure}[bth]
\includegraphics[width=3in]{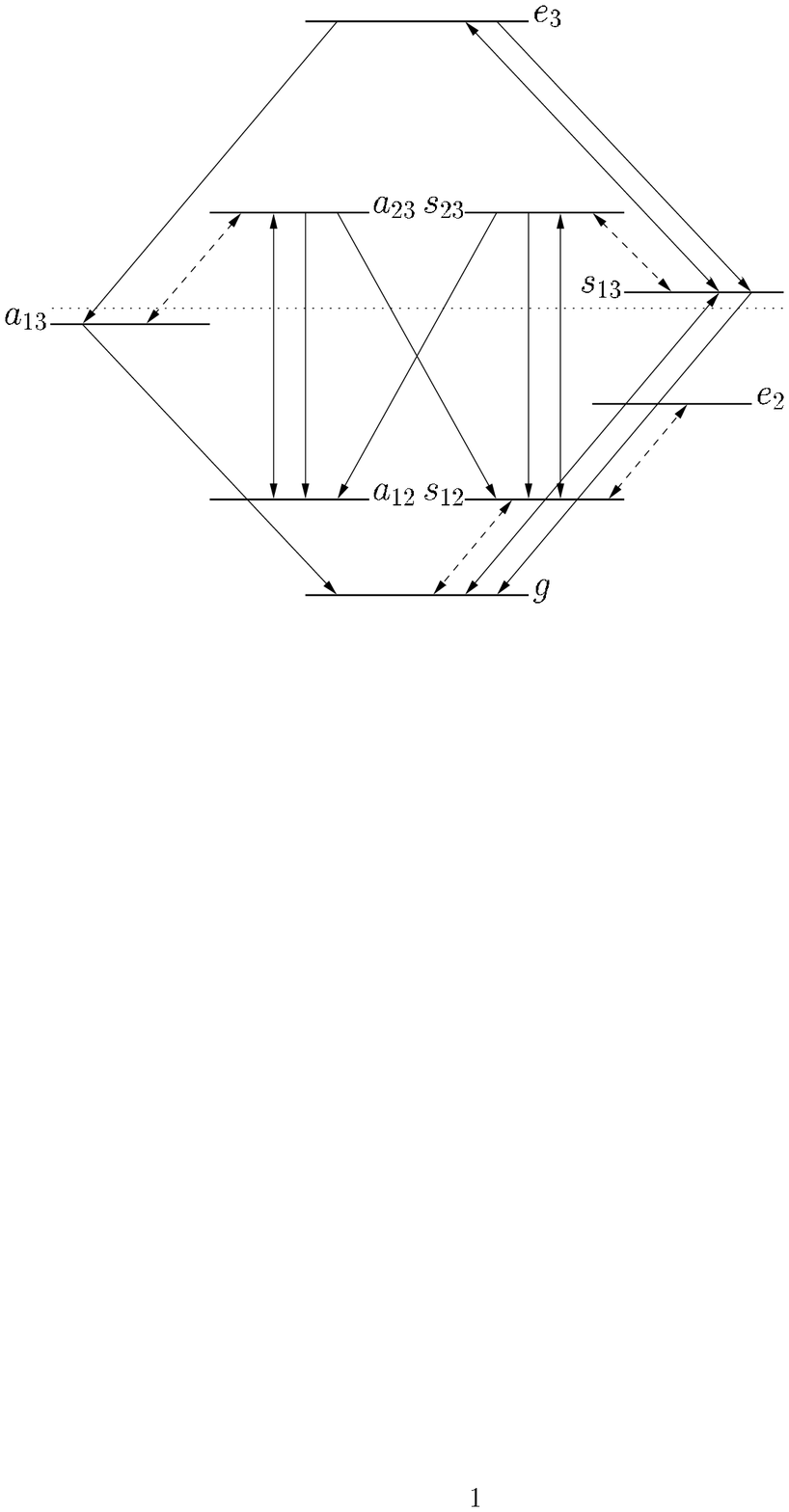}
\vspace{-9cm}
\caption{Dicke states. Simple arrows denote
  decays. Solid and dashed double arrows denote strong
  and weak driving, respectively. 
\label{dicke}}
\end{figure}

We consider two atoms, at a fixed distance {\bf r}, each  a $V$
configuration as shown in 
Fig. \ref{V-system}. 
We assume the 
laser radiation normal to this line, and for the Einstein coefficients,
the  Rabi 
frequencies and the detuning we  assume  the relations
\begin{eqnarray} \label{rel}
\Omega_2 \ll \Omega_3 ~,~~ \Omega_2 \ll \Omega_3^2/A_3,~~ A_2 \approx 0,
~~\Delta_3= 0,
%MD
\end{eqnarray}
$\Delta_2$  arbitrary. The Dicke states are defined as 
\begin{eqnarray} \label{dstates}
|g\rangle &=& |1\rangle|1\rangle,~~~|e_2\rangle = |2\rangle|2\rangle,~~~
|e_3\rangle = |3\rangle|3\rangle\nonumber\\
|s_{jk}\rangle&=& \{|j\rangle|k\rangle+
|j\rangle|k\rangle\nonumber\}/\sqrt{2}\\
i|a_{jk}\rangle&=&\{|j\rangle|k\rangle-
|j\rangle|k\rangle\nonumber\}/\sqrt{2}
\end{eqnarray}
They are symmetric and antisymmetric, respectively, under permutation
of the two atoms. The Dicke states and the possible
transitions are displayed in Fig. \ref{dicke}. 
Solid single and  double arrows indicate decay and strong driving by laser 3,
respectively, while dashed double arrows
indicate the weak driving by laser 2. For $\Omega_2 = 0$, i.e. with
the dashed arrows absent, the 
states decompose into three non-connected
subsets, namely $|e_2 \rangle$, the  four states of the inner ring
and  the four states of the outer ring in
Fig. \ref{dicke}, and the subspaces spanned
by these states will be denoted by dark, inner, and outer subspace,
respectively. As in Ref. \cite{BeHe5}  they  will be associated in the
following with the 
fluorescence periods of intensity 0, 1, and 2: 
\begin{eqnarray}
{\rm dark~state  } & : & |e_2 \rangle\label{dark}\\
{\rm inner~ states~(intensity~1)} &:& |s_{12} \rangle,|s_{2 3} 
\rangle, |a_{1 2}\rangle,  |a_{2 3}\rangle\label{inner}\\
{\rm outer~ states~(intensity~2)}
& : & |g \rangle, |s_{1 3} \rangle, |e_3 \rangle, |a_{1 3}
\rangle\label{outer} 
\end{eqnarray} 
 The weak laser will lead to  slow transitions between the subspaces.

The Bloch equations can be written, with the conditional Hamiltonian
$H_{\rm cond}$ and the reset operation $\cal R$ of
Appendix \ref{A},  in the compact form  \cite{BeHe4,He}
\begin{equation}\label{Bloch}
\dot{\rho} = -~\frac{\rm i}{\hbar}~[H_{{\rm cond}} \rho - \rho
H^\dagger_{{\rm cond}}] + {\cal R}(\rho)~. 
\end{equation}
The operator $H_{\rm cond}$ is of the form
\begin{equation} \label{formhcond}
H_{\rm cond} = H_{\rm cond}^0 + H_{\rm cond}^1(\Omega_2)
\end{equation}
where the operator $H_{\rm cond}^0$ depends on $\Omega_3$ and 
on the dipole-dipole coupling constant $C_3(r)$,
 while $H_{\rm cond}^1$ is linear in $\Omega_2$
and does not depend on $C_3$ and $\Omega_3$.  The super-operator ${\cal
  R}$ depends on $C_3$. 

\subsection{Intensity periods
%jumps modeled as transitions between
and subspaces }
For a {\em single} atom as in Fig. \ref{V-system}, with macroscopic light
and dark periods, the 
stochastic sequence of individual photon emissions can be directly
analyzed by the quantum jump approach
\cite{HeWi,Wi,He,HeSo,MC,QT,PleKni}, using the existence of different
time scales. To high precision it
yields a telegraph process  and the transition 
rates between the periods \cite{BeHe1,BeHeSo} . A more heuristic
approach assumes that 
during a light period the density matrix of the atom lies in the
subspace spanned by $|1\rangle$ and $|3\rangle$ and that during a dark
period the state is given by $|2\rangle$ \cite{Cook}.
One can then use the Bloch equations to calculate the 
build-up, during a time $\Delta t$, of a population outside the
respective subspace and obtains from this the probability of
leaving the subspace. This probability 
is then interpreted as the transition probability from one period to
the other. The results agree with those of the more 
microscopic quantum jump approach \cite{HeWi,Wi,HePle1,HePle2}.

This  idea will be used here for {\em two} dipole-interacting $V$
systems. We associate each of the three types of fluorescence periods
with one of the subspaces spanned by the states in
Eq. (\ref{dark}) - (\ref{outer}) and model transitions between periods as
transitions between the corresponding subspaces. Without
dipole interaction this is the same assumption as for a single
atom, and with the interaction it has been tested numerically in
Ref. \cite{beige} to hold as long as the atomic separation is larger
than a third wavelength of the strong transition.  

Thus, at a particular time $t_0$,
the  density matrix $\rho(t_0)$ of the two atoms is assumed to lie in
one of the subspaces.  
 Then, during a short time $\Delta t$, satisfying 
\begin{equation}
\label{4.1} 
\Omega_3^{-1}\!,A_3^{-1}\ll\Delta t\ll\Omega_2^{-1},
\end{equation}
the system will go over to a density matrix $\rho(t_0+\Delta t)$ which
contains small populations in the other subspaces, due to the driving by $\Omega_2\!\neq \!0$.
The time derivatives of these populations at $t_0 + \Delta t$  give the
transition rates to  these subspaces because, as shown in Appendix B, they
are independent of the particular 
choice of $\Delta t$ and of the particular density matrix
$\rho(t_0)$, as long as Eq.~(\ref{4.1}) is fulfilled.
 These rates can be interpreted   
as transition rates between corresponding intensity periods, just as in the
one-atom case.  

A straightforward calculation using 
Eq.~(\ref{Bloch}) yields  the exact relations
%MD
\begin{eqnarray}
\label{4.2}
\lefteqn{
\nonumber\frac{\rm d}{\rm dt}\sum_{\rm outer}\langle{\rm
  outer}|\rho|{\rm outer}\rangle}
\\&&\hspace{1ex}=\Omega_2 {\rm Im} \Big\{\sqrt{2}\langle s_{12}|\rho|g\rangle
+\langle s_{23}|\rho|s_{13}\rangle+\langle
a_{23}|\rho|a_{13}\rangle\Big\}
\end{eqnarray}
\begin{eqnarray}
\label{4.3}
\frac{\rm d}{\rm dt}\langle e_2|\rho|e_2\rangle=\sqrt{2}\Omega_2 {\rm Im} 
\langle s_{12}|\rho|e_2\rangle
\end{eqnarray}
\begin{eqnarray}
\label{4.4}
\lefteqn{
\frac{\rm d}{\rm dt}\sum_{\rm inner}\langle{\rm inner}|\rho|{\rm
  inner}\rangle}\nonumber\\&&\hspace{7ex}=-\frac{\rm d}{\rm dt}\Big\{\langle
e_2|\rho|e_2\rangle+ 
\sum_{\rm outer}\langle{\rm outer}|\rho|{\rm outer}\rangle\Big\}
\end{eqnarray}
where $|{\rm   outer}\rangle$ stands for  
$|g\rangle$, $|s_{13}\rangle$, $|e_3\rangle$,
$|a_{13}\rangle$ and   $|{\rm inner}\rangle$  for $|s_{1 2} \rangle,
|s_{2 3} \rangle,|a_{1 2} \rangle, |a_{2 3}\rangle$. 
Thus one has to calculate the coherences on the right-hand side at
time $t_0 + \Delta t$ to {\em first} order in
$\Omega_2 $, with the appropriate initial condition at time $t_0$, to
obtain the 
transition rate  to {\em second} order in $\Omega_2$.  

If $\rho(t_0)$ lies in one of the above
subspaces then by time $t_0+\Delta t$ the system has reached a
quasi-stationary state satisfying 
\begin{equation}\label{decisive}
{\dot \rho}(t_0+\Delta t)= 0 ~~~~~{\rm to~first ~order~in~}\Omega_2,
\end{equation}
as shown in Appendix B. This is also true  for a single atom and is
the decisive equation.

To obtain from this the coherences  to first order in $\Omega_2$  we write
$$
\rho(t_0+\Delta t) = \rho^0 + \rho^1 + \cdots
$$
where $\rho^k$ is of order  $\Omega_2^k$. 
Putting $\dot \rho=0$ in Eq. (\ref{Bloch}) and inserting the expansion
for $\rho$ one obtains in zeroth order
\begin{eqnarray} \label{mast2}
0 &=& - {{\rm i} \over \hbar} \left[
H_{\rm cond}^0 \rho^0 - \rho^0 H_{\rm cond}^{0\,\dagger} \right] 
+{\cal R} (\rho^0) 
\end{eqnarray}
and in first order in $\Omega_2$
%MD
\begin{eqnarray} 
\label{mast3}
\lefteqn{
0 = - {{\rm i} \over \hbar} \left[
H_{\rm cond}^0 \rho^1 - \rho^1 H_{\rm cond}^{0\, \dagger} 
+ H_{\rm cond}^1 \rho^0 - \rho^0 H_{\rm cond}^{1\,\dagger} \right]}
\nonumber\\&&\hspace{39ex}+{\cal R} (\rho^1). 
\end{eqnarray}
Thus $\rho^0$ is an
equilibrium state  for $\Omega_2=0$, taken to lie in the appropriate
subspace. For the dark state and  the subspace spanned by the inner
states one has 
\begin{eqnarray}
\label{ss0}
\rho^0\equiv \rho^0_{\rm dark}&=& |e_2\rangle\langle e_2|
\end{eqnarray}
\begin{eqnarray}
\label{ss1}
\lefteqn{
\rho^0\equiv \rho^0_{\rm inner}=\frac{1}{2}\big\{\rho_{{\rm
    ss}}^{(A)}\otimes|2\rangle\langle 
2|+|2\rangle\langle 2|\otimes\rho_{{\rm ss}}^{(B)}\big\}}\nonumber\\
\nonumber&&\hspace{12ex}=\frac{1}{4}\frac{A_3^2+\Omega_3^2}{A_3^2+2\Omega_3^2}
\{|s_{12}\rangle\langle s_{12}|+|a_{12}\rangle\langle a_{12}|\}\\
\nonumber&&\hspace{12ex}+\frac{1}{4}\frac{\Omega_3^2}{A_3^2+2\Omega_3^2}
\{|s_{23}\rangle\langle s_{23}|+|a_{23}\rangle\langle a_{23}|\}\\
&&\hspace{3ex}+\frac{i}{2}\frac{\Omega_3 A_3}{A_3^2+2\Omega_3^2}
\{|s_{12}\rangle\langle s_{23}|-|a_{12}\rangle\langle a_{23}|\}
+{\rm H.c.}
\end{eqnarray}
by symmetry, independently of $C_3$, where $\rho_{{\rm ss}}^{(A,B)}$
are the steady states of the 
individual atoms in the $1$-$3$ subspace (for $\Omega_2=0$ and $C_3=0$).
For the subspace spanned by the outer states one calculates 
\begin{eqnarray}
\label{ss2}
\lefteqn{
\rho^0 \equiv \rho^0_{\rm
  outer}\propto\Big[\big\{(A_3^{2}+\Omega_3^{2})^2+A_3^{2}\,|C_3|^{2}+2A_3^{3}\,{\rm 
      Re}\,C_3\big\}|g\rangle\langle g |}\nonumber
\\\nonumber&&\hspace{4ex}+  \big\{i\sqrt {2}\,A_3\,\Omega_3\,
  (A_3^{2}+\Omega_3^{4}+A_3\,C_3 )  |g\rangle\langle 
s_{13}| + {\rm H.c.} \big\}\\
\nonumber&&\hspace{4ex}-\big\{ A_3\,\Omega_3^{2}\, (
A_3+C_3 ) |g\rangle\langle e_3| + {\rm H.c.}\big\}\\ 
\nonumber&&\hspace{4ex}+ \Omega_3^{2}\,
(2 A_3^{2}+\Omega_3^{2} ) 
|s_{13}\rangle\langle s_{13}|+\Omega_3^{4}\,\big\{|e_3\rangle\langle e_3|
+ |a_{13}\rangle\langle a_{13}|   \big\}\\
&&\hspace{4ex}+\big\{i\sqrt {2}\,A_3\,\Omega_3^{3}
|s_{13}\rangle\langle e_3| + {\rm H.c.}\big\}\Big]
\end{eqnarray}
One checks that for $C_3=0$ this becomes $\rho^0_{\rm outer}\propto
\rho_{{\rm  ss}}^{(A)}\otimes\rho_{{\rm  ss}}^{(B)}$, the expression 
for two independent atoms.

We will denote the transition rates between the subspaces by
$p_{ij}$. Here 
$i,j = 0,1,2$ refer  to the dark, inner and outer subspace, respectively,
(and thus to the corresponding intensities). The $p_{ij}$ will be 
determined  to second order in $\Omega_2$. As expected, $p_{02}$ and
$p_{20}$ will turn out to be zero.
\subsection{Calculation of $p_{12}$}
We start from  $\rho^0=\rho^0_{\rm inner}$ in Eq. (\ref{ss1}) as
initial state.  
For the the transition rate $p_{12}$ to  the outer subspace one needs, in
view of Eq. (\ref{4.2}), three coherences of 
 $\rho^1$ between the inner and outer subspace. To obtain these we 
write $\{|x_i\rangle\} = \{|s_{12} \rangle, |s_{2 3} 
\rangle,  |a_{1 2}
\rangle, |a_{2 3}\rangle\}$ (inner states) and $\{|y_j\rangle\} = \{|g
\rangle, |s_{1 3} \rangle,  |e_3 \rangle, |a_{1 3} \rangle\}$ (outer 
states) for the corresponding bases and decompose
\begin{eqnarray} \label{GS2}
\rho^1 &=& \sum_{i,j} \rho_{ij}^1 \, |x_i \rangle \langle
y_j|
+ \rho_{ij}^{1*} \, |y_j \rangle \langle x_i| + {\rm other~terms}~.
\end{eqnarray}
Inserting this into Eq. (\ref{mast3}) and taking matrix elements
with   $\langle x_{i_0}|$ on the left and $|y_{j_0}\rangle$ on the right
gives 
\begin{eqnarray} \label{system}
0 &=& \frac{i}{\hbar}\langle x_{i_0}| \rho^0_{\rm inner} H_{\rm cond}^{1 \, \dagger} |y_{j_0}\rangle
- \frac{i}{\hbar}\sum_i \rho_{ij_0}^1 \, \langle x_{i_0}| H_{\rm cond}^0 |x_i
\rangle\nonumber\\  
&&+ \frac{i}{\hbar} \sum_j \rho_{i_0j}^1 \, \langle y_j|H_{\rm cond}^{0\, \dagger} |y_{j_0} \rangle  
\nonumber \\
& & + \sum_{i,j} (A_3+{\rm Re} \, C_3) \rho^1_{ij} 
\langle x_{i_0}| R_+ |x_i \rangle \langle y_j| R_+^\dagger
|y_{j_0}\rangle \nonumber\\ 
&&+ \sum_{i,j}(A_3-{\rm Re} \, C_3)\rho^1_{ij}  
\langle x_{i_0}| R_- |x_i \rangle \langle y_j| R_-^\dagger |y_{j_0}\rangle ~.
\end{eqnarray}
This is a system of 16 linear equations for the 16 coherences
$\rho^1_{ij}$, of which only three are needed in Eq. (\ref{4.2}). Due
to the symmetry of $H_{\rm cond}$ and $R_+$ and antisymmetry of
$R_-$ under the interchange of the two atoms, the system
decouples. Taking for $|x_{i_0}\rangle$ and $|y_{j_0}\rangle$ either
both symmetric or both antisymmetric states and putting  the eight
coherences into the column vector
\begin{widetext}
\begin{equation}
\label{coh}
\tilde{\bf \rho} \equiv  (\rho_{s_{12}g}^1,~
\rho_{s_{12}s_{13}}^1,~\rho_{s_{12}e_3}^1, ~\rho_{s_{23}g}^1,~
\rho_{s_{23}s_{13}}^1, ~\rho_{s_{23}e_3}^1,~\rho_{a_{12}a_{13}}^1,~
\rho_{a_{23}a_{13}}^1)^T
\end{equation}
\end{widetext}
one obtains the equation
\begin{equation}
 \label{form}
({\bf A}-i\Delta_2{\bf 1})\tilde{\bf \rho}= {\bf a}_1
\end{equation}

\begin{widetext}
\noindent where
\begin{eqnarray}
\lefteqn{{\bf A}=}\\
&&\nonumber\hspace{-2ex}\left [\begin {array}{cccccccc} 
0 & -i\Omega_3/\sqrt{2}& 0 & i\Omega_3/2 & 
-(A_3\!+\!{\rm Re}C_3)/\sqrt{2} & 0 & 0 & ({\rm Re}C_3\!-\!A_3)/\sqrt{2}\\
-i\Omega_3/\sqrt{2} & (A_3\!+\!C_3^\ast)/2 & -i\Omega_3/\sqrt{2} & 0 &
i\Omega_3/2 & -(A_3\!+\!{\rm Re}C_3)/\sqrt{2} & 0 & 0\\
0&-i\Omega_3/\sqrt{2}&A_3&0&0&i\Omega_3/2&0&0\\
i\Omega_3/2&0&0&A_3/2&-i\Omega_3/\sqrt{2}&0&0&0\\
0&i\Omega_3/2&0&-i\Omega_3/\sqrt{2}&(A_3\!+\!C_3^\ast/2)
&-i\Omega_3/\sqrt{2}&0&0\\
0&0&i\Omega_3/2&0&-i\Omega_3/\sqrt{2}&3A_3/2&0&0\\
0&0&0&0&0&-(A_3\!-\!{\rm Re}C_3)/\sqrt{2}&(A_3\!-\!C_3^\ast)/2&-i\Omega_3/2\\
0&0&0&0&0&0&-i\Omega_3/2&(A_3\!-\!C_3^\ast/2)
\end{array}\right ]
\end{eqnarray}
\begin{eqnarray}
{\bf a}_1 =\frac{i\Omega_2\Omega_3}{4(A_3^2+2\Omega_3^2)}\left[
\sqrt{2}\,\frac{\Omega_3^2+A_3^2}{\Omega_3},
i\,A_3,0,-i\sqrt{2}\,A_3,\Omega_3,0,
-i\,A_3,
\Omega_3\right]^T.
\end{eqnarray}
\end{widetext}
Inverting the $8\times8$ matrix ${{\bf A} - i \Delta_2 {\bf 1}}$ by Maple
yields $\tilde{\bf 
  \rho}$ and the coherences. The
result is complicated and not illuminating.  Inserting the required
coherences 
into Eq. (\ref{4.2}) one obtains, to  first order in
Re$\,C_3$ and Im$\,C_3$ and to second order in $\Omega_2$,
\begin{eqnarray}
\label{p12}
\lefteqn{
p_{12}=
\Omega_2^2 \Big\{ \frac{A_3 \Omega_3^2 }{\Omega_3^4-8 \Delta_2^2
\Omega_3^2+4 A_3^2 \Delta_2^2+16 \Delta_2^4 
}}\nonumber \\&&\hspace{-5ex}
+{\rm Re\,}C_3(r)\frac{2 A_3^2
 \Omega_3^2 (\Omega_3^4\!-\!4 A_3^2 \Delta_2^2\!-\!16\Delta_2^4)  
}{(A_3^2\!+\!2 \Omega_3^2) (\Omega_3^4\!-\!8 \Delta_2^2 \Omega_3^2\!+\!4 A_3^2
\Delta_2^2\!+\!16 \Delta_2^4)^2}\Big\}.
\end{eqnarray}
 Note that only Re$\, C_3$ appears and that the  terms linear in Im$\,
 C_3$ have canceled.  
\subsection{Calculation of $p_{10}$}
To determine $p_{10}$ we  use Eq. (\ref{4.3}) and start again from $\rho^0
= \rho^0_{\rm inner}$ as initial condition in Eq. (\ref{mast3}), but
now have to determine $\langle s_{12}|\rho^1|e_2\rangle$. Replacing
$\{|y_i\rangle\}$ by $|e_2\rangle$ and choosing $|x_{i_0}\rangle=
|s_{12}\rangle,~|s_{23}\rangle$ in Eq. (\ref{system}), one obtains  two
inhomogeneous linear equations for $\langle s_{12}|\rho^1|e_2\rangle$
and $\langle s_{23}|\rho^1|e_2\rangle$. These
equations do not depend on $C_3$, since $R_{\pm}$ and
$R_{\pm}^\dagger$ vanish on $|e_2\rangle$ and since $C_3$ does not
appear in the part of $H_{\rm cond}$ acting on the inner
states. Therefore $\langle s_{12}|\rho^1|e_2\rangle$
and $\langle s_{23}|\rho^1|e_2\rangle$ are independent of $C_3$. By a
simple calculation one obtains $\langle s_{12}|\rho^1|e_2\rangle $
and inserting this into Eq. (\ref{4.3}) yields, to second order in $\Omega_2$,
\begin{equation}
\label{p10}
p_{10}=\Omega_2^2\frac{A_3\Omega_3^2(A_3^2+4\Delta_2^2)}{(A_3^2+2\Omega_3^2)
[(\Omega_3^2-4\Delta_2^2)^2+4\Delta_2^2A_3^2]}~.
 \end{equation}  
This is independent of $C_3$ and is the same  as for two
independent atoms, namely the transition rate for a single atom from a
light to a dark period \cite{Wi}. 
\subsection{Calculation of $p_{01}$}
To determine $p_{01}$ we  use Eq. (\ref{4.4}). One also needs
$\langle s_{12}|\rho^1|e_2\rangle$, as seen from Eq. (\ref{4.3}), but
in this case  one has to  start 
from $\rho^0 = \rho^0_{\rm dark}$ as initial condition in
Eq. (\ref{mast3}). Therefore one obtains the same equations for
$\langle s_{12}|\rho^1|e_2\rangle$
and $\langle s_{23}|\rho^1|e_2\rangle$ as before, except for the
inhomogeneous part. One  has independence of $C_3$ and easily
solves for $\langle s_{12}|\rho^1|e_2\rangle$.
 
For the remaining coherences   needed in Eq. (\ref{4.4}),
i.e. those in Eq. (\ref{4.2}), one obtains the same form as in
Eq. (\ref{form}), with the same ${\bf A}$, but now with ${\bf a}_1 = 0$
since the term containing $\rho^0$ vanishes. Therefore these
coherences vanish here and hence $p_{02}=0$. Physically this means
that in our formulation of the problem there are no direct transitions
from a dark period to a period of intensity 2. 

From Eq. (\ref{4.4}) one now
obtains, to second order in $\Omega_2$,
\begin{equation} \label{p01}
p_{01}=2\Omega_2^2\frac{A_3 \Omega_3^2}
{(\Omega_3^2-4\Delta_2^2)^2+4\Delta_2^2A_3^2}
 \end{equation}
This is  independent of $C_3$ and is the same  as for two
independent atoms, namely twice the transition rate for a single atom from a
dark to a light period.
\subsection{Calculation of $p_{21}$}
The transition rate $p_{21}$ is obtained from Eq. (\ref{4.4}) and the
required coherences are again those appearing in Eq. (\ref{4.2}) and
(\ref{4.3}), now with $\rho^0 = \rho^0_{\rm outer}$ 
as initial condition. For $\langle s_{12}|\rho^1|e_2\rangle$
and $\langle s_{23}|\rho^1|e_2\rangle$ one obtains the same two equations
as before, except for the inhomogeneous part which now vanishes. Hence 
these two coherences vanish now and as a consequence
$p_{20}=0$. Physically this means 
that in our formulation there are no direct transitions from a period
of intensity 2 to a dark period. For the coherences in Eq. (\ref{coh})
one has the same equation as Eq. (\ref{form}), with the same matrix
${\bf A}$ but with ${\bf a}_1$ replaced by 
\begin{widetext}
\begin{eqnarray}
{\bf a}_2=-i\Omega_2\Big/\big\{\sqrt{2}\left
      (4\,{\Omega_3}^{4}+4\,{\Omega_3}^{2}{{A_3}}^{2}+
{A_3}^{2}{{{\rm Re}\,C_3}}^{2}+2\,{A_3}^{3}{{\rm Re}\,C_3}+{{A_3}}^{2}
{{{\rm Im}\,C_3}}^{2}+{A_3}^{4}\right )\big\}\\\times
\left [\begin {array}{c} {{\Omega_3}}^{4}+2\,{{\Omega_3}}^{2}{{A_3}}^
{2}+{{A_3}}^{2}{{\rm Re}\,C_3}^{2}+2\,{{A_3}}^{3}{\rm Re}\,C_3+{{A_3}}^{2}
{{{\rm Im}\,C_3}}^{2}+{{A_3}}^{4}\\\noalign{\medskip}i{
\Omega_3}\,\sqrt {2}{A_3}\,\left ({{A_3}}^{2}+{A_3}
\,{\rm Re}\,C_3+i{\rm Im}\,C_3\,{A_3}+{{\Omega_3}}^{2}\right )
\\\noalign{\medskip}-{{\Omega_3}}^{2}\left ({A_3}+{\rm Re}\,C_3+
i{\rm Im}\,C_3\right ){A_3}\\\noalign{\medskip}i{\Omega_3}\,
{A_3}\,\left (-{{\Omega_3}}^{2}-{{A_3}}^{2}-{A_3}\,{\rm Re}\,C_3+
i{\rm Im}\,C_3\,{A_3}\right )\\\noalign{\medskip}
{{\Omega_3}}^{2}\left ({{\Omega_3}}^{2}+2\,{{A_3}}^{2}\right )/\sqrt {2}
\\\noalign{\medskip}i{{\Omega_3}}^{3}{A_3}
\\\noalign{\medskip}0\\\noalign{\medskip}{{\Omega_3}}^{4}/\sqrt {2}
\end {array}\right ]
\end{eqnarray}
\end{widetext}
Inserting the resulting coherences into Eq. (\ref{4.4}) gives, to
first order in Re$\,C_3$ and ${\rm Im}\,C_3$ and to second order in $\Omega_2$,
\begin{widetext}
\begin{eqnarray}
\label{p21}
\lefteqn{
p_{21}= \Omega_2^2\Big\{\frac{2A_3\Omega_3^2 (A_3^2+4
\Delta_2^2)}
{(\Omega_3^4-8 \Delta_2^2 \Omega_3^2+16 \Delta_2^4
+4 A_3^2 \Delta_2^2) (A_3^2+2 \Omega_3^2)}}
\nonumber \\&&\hspace{1ex}+{\rm Re\,}C_3(r)\,\frac{4 A_3^2 \Omega_3^2    (A_3^4\Omega_3^4
+4 A_3^2 \Omega_3^6-12 A_3^2 \Delta_2^2
\Omega_3^4-64 A_3^2 \Delta_2^6-4 A_3^6 \Delta_2^2
-32 A_3^4\Delta_2^4-64 \Delta_2^4 \Omega_3^4+16 \Delta_2^2 \Omega_3^6
)}{(A_3^2+2 \Omega_3^2)^3 (\Omega_3^4-8 \Delta_2^2
\Omega_3^2+4 A_3^2 \Delta_2^2+16 \Delta_2^4)^2}\Big\}
\end{eqnarray}
\end{widetext}
where again the terms containing Im$\, C_3$ have canceled.

\subsection{Discussion.} 
\begin{figure}[bth]
\hspace{-2cm}\includegraphics[width=4in]{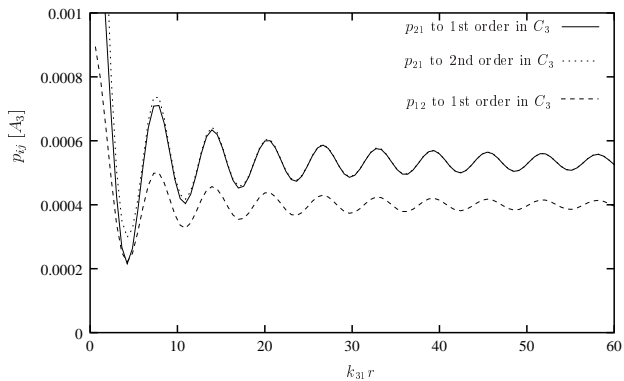}
\caption{Transition probabilities $p_{21}$ to first and second order
  in $C_3$  and $p_{21}$ to first order, for
  $\Omega_3\!=\!0.5A_3$,  
$\Omega_2\!=\!0.01A_3$, zero detuning. The 
contribution to $p_{21}$ arising from the second order in $C_3$ is
small.
\label{p2112}} 
\end{figure}

If  one computes the coherences in Eqs. (\ref{4.2})
and (\ref{4.4}) to second order in $C_3$ one obtains
$p_{12}$ and $p_{21}$ to second order in $\,C_3$. 
 The resulting expressions are not enlightening and therefore 
not given here, but they do depend on $({\rm Im}\,C_3)^2$.
Fig. \ref{p2112} 
shows how small the second-order
dipole-dipole contribution to $p_{21}$ is for the parameters of the simulations
and for distances larger than half a 
wave length. 
For smaller distances the 
results are probably not applicable anyway, as discussed in
Ref. \cite{BeHe5}.  

For $\Delta_2=0$ the rates $p_{12}$ and $p_{21}$ simplify to
\begin{eqnarray}
\label{p12Delta0}
p_{12}&=& \Omega_2^2 \Big\{ \frac{A_3}{\Omega_3^2} + {\rm Re\,}C_3(r)\, 
\frac{2 A_3^2 
}{\Omega_3^2 (A_3^2+2 \Omega_3^2)}\Big\}
\end{eqnarray}
\begin{eqnarray}
\lefteqn{
p_{21}=\Omega_2^2 \Big\{\frac{2A_3^3}{\Omega_3^2(A_3^2\!+\!2 \Omega_3^2) } 
+ }\nonumber\\&&\hspace{20ex}{\rm Re\,}C_3(r)\, \frac{4 
  A_3^4 (A_3^2\!+\!4 \Omega_3^2)}{\Omega_3^2 (A_3^2\!+\!2 
\Omega_3^2)^3}\Big\}\label{p21Delta0}
\end{eqnarray}
and  one sees that the coefficients of the Re$\,C_3$ term in
Eqs. (\ref{p12Delta0}) and 
(\ref{p21Delta0})  are positive. For $\Delta_2=0$, therefore, $p_{12}$ and 
$p_{21}$ vary with the atomic distance  {\em in phase} with Re$\,C_3$. 
For $\Delta_2 \ne 0$, however, the coefficients of Re$\,C_3$ in Eqs.
(\ref{p12}) or (\ref{p21}) can become zero or negative. In the first
case $p_{12}$ or $p_{21}$ become constant in $r$, while in the second
case they vary {\em opposite} in phase to Re$\,C_3$. 

It will be shown in the next sections that this dependence of $p_{12}$
and $p_{21}$ on the detuning of the weak laser entails a
corresponding behavior of the double jump rate and an
opposite behavior of the mean durations $T_1$ and $T_2$. This opposite
behavior of $T_1$ and $T_2$ is easy to understand since they are
related to the inverse of the transition rates.  

\section{Double jumps: Comparison of simulations with theory}\label{DJTheory}
A double jump is defined as a transition from a double-intensity period to dark
period, or vice versa, within a prescribed  time interval $\Delta
T_{\rm DJ}$.  Now, to distinguish different periods in experiments and in
simulations one has to use an average photon intensity, 
obtained e.g. by means of averaging over a time window. This window has to
be large enough to contain enough emissions, but must not be too large
in order not to overlook too many short periods. 
Our simulations employ a procedure similar to that in Ref. \cite{BeHe5}
and use a moving window \cite{window} of fixed 
width, denoted by $\Delta T_{\rm w}$. The time interval $\Delta
T_{\rm DJ}$ should be larger than $\Delta T_{\rm w}$.

We consider the fluorescence periods as a
telegraph process with three steps and use the $p_{ij}$ of  the last
section as transition rates. At first the influence
of the averaging window $T_{\rm w}$ will be neglected. 

\begin{figure}[bth]
\hspace{-2cm}\includegraphics[width=4in]{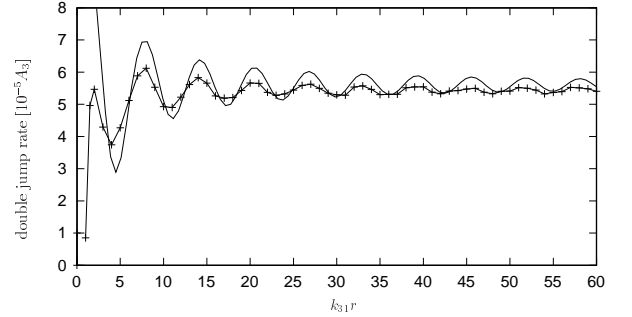}
\caption{Double jump rates. ~Simulation~~ $+++$~,
~theory~~  $-\!\!\!\!-\!\!\!-\!\!\!-$~~  
~uncorrected for averaging window 
($\Omega_3\!=\!0.5~A_3$, $\Omega_2\!=\!0.01~A_3$, zero
detuning).
\label{DSVergleichunkorrneu}}  
\end{figure}

The rate of {\em   downward} double jumps is  obtained 
as follows. For $ i$ = 0, 1, 2, let $n_i$ be the mean number of periods
of intensity $i$ per unit time. For a long path of length $T$ the total number of
periods of intensity $i$ is then $N_i(T) = n_i T$. At the end of each period of intensity 2
there begins a period of intensity 1, and the probability for this period
of intensity 1 to be shorter than $\Delta T_{\rm DJ}$   is given by 
$$
1-\exp\{-(p_{10}+p_{12}) \Delta T_{\rm DJ}\}.
$$
At the end of a period of intensity 1  the branching
ratio for a transition to a period of intensity 0 is $p_{10}/(p_{10}+p_{12})$.
Thus during time $T$ the total number of such
downward double jumps, denoted by $N^{20}_{\rm DJ}(T)$, is 
$$
N^{20}_{\rm DJ}(T)\!=\! N_2(T)
\frac{p_{10}}{(p_{10}\!+\!p_{12})}\Big\{1\!-\exp\{-(p_{10}\!+\!p_{12}) \Delta
T_{\rm DJ}\}\Big\} 
$$
and therefore the rate, $n^{20}_{\rm DJ}$, of downward double jumps
within $\Delta T_{\rm DJ}$ is
\begin{equation} \label{5.1}
n^{20}_{\rm DJ}\!=\! n_2
\frac{p_{10}}{(p_{10}\!+\!p_{12})}\Big\{1\!-\exp\{-(p_{10}+p_{12})
  \Delta T_{\rm DJ}\}\Big\}.  
\end{equation}
In a similar way one finds that the rate, $n^{02}_{\rm DJ}$, of upward
double jumps within $\Delta T_{\rm DJ}$ is 
\begin{equation} \label{5.2}
n^{02}_{\rm DJ}\!=\! n_0
\frac{p_{12}}{(p_{10}\!+\!p_{12})}\Big\{1\!-\exp\{-(p_{10}+p_{12}) 
  \Delta T_{\rm DJ}\}\Big\}.  
\end{equation}
It remains to determine $n_0$ and $n_2$. Since a period of intensity 1 ends
with a transition to a period of either intensity 0 or intensity 2 one has,
with the respective branching ratios, 
\begin{equation} \label{5.3}
n_0=\frac{p_{10}}{p_{10}+p_{12}}n_1
\end{equation}
\begin{equation} \label{5.4}
n_2=\frac{p_{12}}{p_{10}+p_{12}}n_1.
\end{equation}
If one denotes by $T_i$ the mean durations of a period of intensity $i$,
one has 
\begin{equation} \label{5.5}
\sum_{i=0}^2 n_iT_i = 1.
\end{equation}
Moreover, one has 
\begin{equation} \label{5.6}
T_0 =1/p_{01},~~~~~ T_1=1/(p_{10}+p_{12}),~~~~~ T_2=1/p_{21}
\end{equation}
and this then gives 
\begin{equation} \label{5.7}
n_0= \frac{p_{01}p_{21}}{p_{01}p_{21}+p_{21}p_{10} + p_{01}p_{12}} p_{10}
\end{equation}
\begin{equation} \label{5.8}
n_2 = \frac{p_{01}p_{21}}{p_{01}p_{21}+p_{21}p_{10} + p_{01}p_{12}} p_{21}.
\end{equation}
From this, together with Eqs. (\ref{5.1}) and (\ref{5.2}), one sees
immediately that the rates of upward and downward double jumps are
equal,
\begin{equation} \label{5.9}
n^{02}_{\rm DJ}=n^{20}_{\rm DJ}.
\end{equation}
This fact was also observed in the simulations. The combined number of
double jumps therefore equals  
\begin{eqnarray} 
\label{5.10}
\lefteqn{n_{\rm DJ}\equiv n^{02}_{\rm DJ}+n^{20}_{\rm DJ}} \nonumber
\\&&\hspace{5ex}= 
2 \frac{p_{01}p_{10}p_{12}p_{21}}{(p_{01}p_{21}+p_{21}p_{10} +
p_{01}p_{12})(p_{01}+p_{12})}\nonumber
\\&&\hspace{13ex}\times
\Big\{1-\exp\{-(p_{10}+p_{12})
\Delta T_{\rm DJ}\}\Big\}. 
\end{eqnarray}
For $\Delta T_{\rm DJ}\ll T_1$ and by expanding the exponential, this
gives for the combined double jump rate, without correction for the
averaging window,
\begin{equation} \label{5.11}
n_{\rm DJ} = 
2 \frac{p_{01}p_{10}p_{12}p_{21}}{p_{01}p_{21}+p_{21}p_{10} +
  p_{01}p_{12}}  \Delta T_{\rm DJ}.
\end{equation}
Fig. \ref{DSVergleichunkorrneu}  
shows a comparison of this result with 
data from the simulations. 
Except for atomic distances  less
than about three quarters of the wave length of the strong transition
the agreement appears as quite reasonable, and the disagreement for
small distances is 
not unexpected since there the intensities start to decrease and a
description by a telegraph process may be no longer a good
approximation, as pointed out in Ref. \cite{BeHe5}. But 
one observes that the theoretical result is systematically
above the simulated curve.  This  seeming disagreement, however, is
easily explained and can be taken care of as follows.
\subsection{Corrections for averaging window}
\begin{figure}[bth]
\hspace{-2cm}\includegraphics[width=4in]{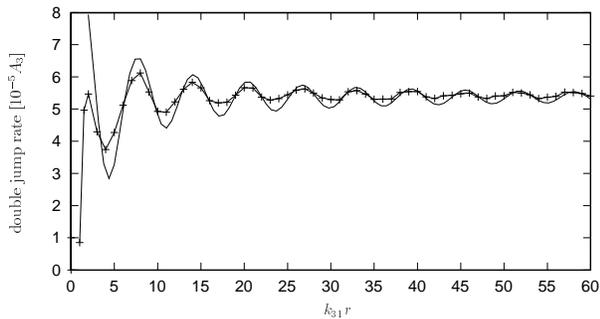} 
\caption{As in Fig \protect\ref{DSVergleichunkorrneu}
, but theory
  corrected for averaging window.
\label{DSVergleich70neu}}
\end{figure}

 We recall that the simulated data
were obtained by averaging the numerical photon emission  times with a
moving window of length $\Delta T_{\rm w}$. Then, roughly, periods
which are shorter than about two thirds of the window length are overlooked,
and therefore the number of recorded (or observed) periods of
type 2,  which enters Eq. (\ref{5.1}), is smaller than that given by
Eq. (\ref{5.8}). The recorded or observed number is denoted by
$n_{2,\rm cor}$. It is  approximately given by
\begin{equation} \label{5.12}
n_{2,\rm cor} = n_2 \exp\{-p_{21}\frac{2}{3}\Delta T_{\rm w}\},
\end{equation}
and this  expression should be inserted into Eq. (\ref{5.1}) for $n_2$.
In this way one obtains the corrected theoretical curve in
Fig. \ref{DSVergleich70neu}. 
The curve changes very little if instead
of two thirds one takes 60\% or 70\% of $\Delta
T_{\rm w}$. It is seen that the 
agreement with the simulated data is much improved for 
distances greater than three quarters of a wave length of the strong
transition. 

It still appears, however, that the oscillation amplitudes of the  theoretical
curve are somewhat larger than those of the simulated curve. This is again
understandable as an effect of the averaging procedure. In the simulations
it was noticed numerically that the $r$ dependence of the double jump rate  depended
somewhat on the length of the averaging window $T_{\rm w}$ and  
 distinct features tended  to be somewhat washed out for
 larger $\Delta T_{\rm w}$, in particular the oscillation amplitudes
 of the simulated 
data decreased with the length of the averaging window. A larger
$\Delta T_{\rm w}$ gave a 
smoother intensity curve, but made the determination of the
transition times between different periods more difficult, while a
shorter averaging window introduced more noise. We found the use
 of $\Delta T_{\rm w} = 114~ A_3^{-1}$ to be a good compromise.
  If it were
possible to choose smaller averaging window  the amplitudes should
increase,  as predicted by the theory. 
\subsection{Detuning}
\begin{figure}[bth]
\hspace{-3cm}\includegraphics[width=4.5in]{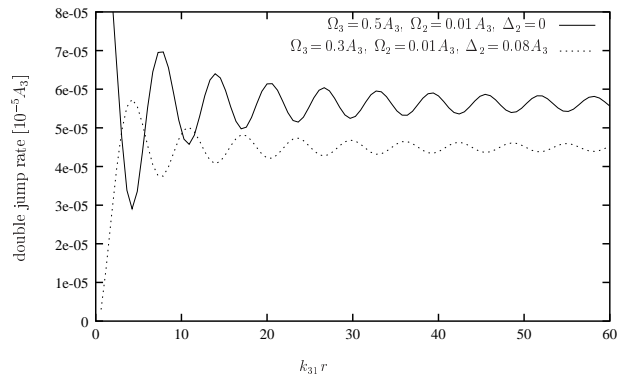}
\caption{Changed oscillatory behavior of the double jump rate for
  increased detuning of the weak driving (uncorrected for averaging
  window).
\label{DSDelta}} 
\end{figure}

One can explicitly insert the expressions for $p_{ij}$ of the last
section into Eq. (\ref{5.11}), but the result becomes unwieldy. One
can show that in an expansion of Eq. (\ref{5.11}) with respect to
Re$\, C_3$  to first order 
 the coefficient of Re$\, C_3$ is positive for 
zero detuning. This
implies that the double jump rate is {\em in phase} with Re$\, C_3(r)$
for the atomic distances under consideration and for zero detuning. 
For increasing detuning the double jump rate can become constant in
$r$ and then change its oscillatory behavior to that of
$-$Re$\,C_3$. An example for the latter  is shown in
Fig. \ref{DSDelta}.   
\section{Duration of fluorescence periods: Effect of averaging
  window }\label{duration} 

The mean durations, $T_0$, $T_1$, and $T_2$, of the three periods were
investigated for cooperative effects in Ref. \cite{BeHe5} by
simulations with averaging windows at discrete times. Here we have
performed similar simulations with a 
moving averaging window. It turns out that both the present and the
previous simulation for $T_i$ are
about 15\% higher than those predicted by Eq. (\ref{5.6}), using the
expressions for $p_{ij}$  of Section \ref{pij} and without correcting for
the use of the averaging window due to which short periods are not
recorded. We will now show how 
this can be taken into account in the theory.

\begin{figure}[bth]
\hspace{-2cm}\includegraphics[width=4in]{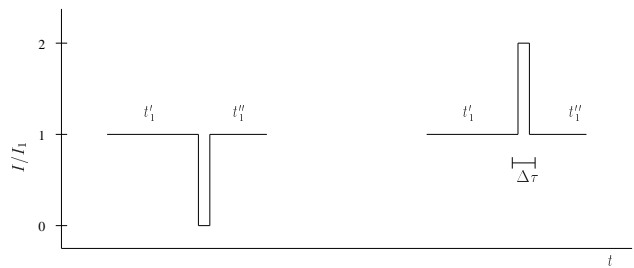}
\caption{If periods of length  $\Delta\tau$ or less are overlooked 
then the distribution of the periods is changed.
\label{rec}}
\end{figure}

 As in Section \ref{DJTheory} we consider a three-step
telegraph process with periods of type 0, 1, and 2,  whose mean
durations are denoted by
$T_0$, $T_1$, and $T_2$, respectively. We assume that periods of length $\Delta
\tau$ or less are not recorded. Fig. \ref{rec} shows periods of type
1 which are interrupted by a short period of  type 0 and 2,
respectively. 
If the respective short periods are not recorded, then
the {\em two}
periods of type 1 in the left part  of the figure are  recorded as a
{\em single} longer period, and similarly for the right part of the
figure. This leads to an apparent decrease of shorter periods of type
1 and to a corresponding increase of longer periods. 

To make this quantitative we put $\lambda_i \equiv 1/T_i$ and denote
the number per unit time of periods of type $i$, whose duration is
less than  $\Delta \tau$, by $n_i^{\Delta \tau}$, i.e.
\begin{equation} \label{6.1}
n_i^{\Delta \tau} = n_i \Big\{1-\exp\{-\lambda_i \Delta \tau\}\Big\}.
\end{equation}
Per unit time, one has $n_0^{\Delta \tau}$ occurrences of the
situation in the left part of Fig. \ref{rec} and $n_2^{\Delta \tau}$
occurrences of 
that in the right part. The probability for one of the periods of type 1
in the left or right part of Fig.  \ref{rec} to have a length  lying
in the time interval $(t_1,t_1+dt_1)$ is 
$2\lambda_1\exp\{-\lambda_1 t_1\}dt_1$, where the factor of 2 comes from
the two possible situations. Therefore, the recorded number, per unit time,
of periods of type 1 with duration in $(t_1,t_1+dt_1)$  is changed
(decreased) by  
\begin{equation} \label{6.2}
2(n_0^{\Delta \tau} + n_2^{\Delta \tau})\lambda_1\exp\{-\lambda_1 t_1\}dt_1.
\end{equation} 
Similarly, the apparent increase of the number, per unit time, of
periods of type 1 with duration in $(t_1,t_1+dt_1)$  is, by Fig. \ref{rec}, 
\begin{eqnarray} 
\label{6.3}
\lefteqn{
(n_0^{\Delta \tau} + n_2^{\Delta \tau})
~~~~~~\int\!\!\!\!\!\!\!\!\!\!\int\limits_{\hspace{-3ex}t_1\le t_1^{\prime} +
  t_1^{\prime\prime}\le t_1+dt_1} dt_1^{\prime} dt_1^{\prime\prime}}
\nonumber\\&&\hspace{16ex}\times
\lambda_1\exp\{-\lambda_1 t_1^{\prime}\}\lambda_1\exp\{-\lambda_1
t_1^{\prime\prime}\} \nonumber   \\
&&\hspace{12ex}= (n_0^{\Delta \tau} + n_2^{\Delta \tau})\lambda_1^2t_1
\exp\{-\lambda_1t_1\}dt_1. 
\end{eqnarray}
Denoting by $\nu_{1\rm rec}(t_1)dt_1$ the actually recorded number, per
unit time, of periods of type 1 with duration in $(t_1,t_1+dt_1)$ one
obtains from the two previous expressions 
\begin{eqnarray} 
\label{6.4}
\lefteqn{
\nu_{1\rm rec}(t_1)dt_1= n_1\lambda_1\exp\{-\lambda_1 t_1\}dt_1}
\nonumber \\ && \hspace{4ex}
+(n_0^{\Delta \tau} + n_2^{\Delta \tau})(\lambda_1^2t_1 -2\lambda_1)
\exp\{-\lambda_1t_1\}dt_1. 
\end{eqnarray}
The average duration of the recorded periods of type 1 will be denoted
by $T_{1,\rm cor}$, and it is given by 
\begin{equation} \label{6.5}
T_{1,\rm cor} = \int_{\Delta \tau}^\infty dt_1\, t_1 \nu_{1\rm rec}(t_1)\bigg/
\int_{\Delta \tau}^\infty dt_1 \nu_{1\rm rec}(t_1).
\end{equation} 
Using Eq. (\ref{6.4}) for $\nu_{1\rm rec}(t_1)$ one obtains, after
an elementary 
calculation and for $\Delta \tau$ satisfying $\Delta \tau/T_1 \ll 1$,  
\begin{equation} \label{6.6}
T_{1,\rm cor} = \frac{1}{p_{10}+p_{12}} +\Delta \tau\left\{1 +
\frac{p_{01}p_{10}+
p_{12}p_{21}}{(p_{10}+p_{12})^2}\right\} .
\end{equation}
The first term is the ideal theoretical value, $T_1$, and the remainder is the
correction due to non-recorded short periods.
In a similar way one obtains
\begin{equation} \label{6.7}
T_{0,\rm cor} = \frac{1}{p_{01}} +\Delta \tau\left\{1 +
\frac{p_{10}}{p_{01}}\right\}
\end{equation}
\begin{equation} \label{6.8}
T_{2,\rm cor} = \frac{1}{p_{21}} +\Delta \tau\left\{1 +
\frac{p_{12}}{p_{21}}\right\} 
\end{equation}
where again the respective first terms are the ideal values, $T_0$ and
$T_2$. 

\begin{figure}[bth]
\hspace{-2cm}\includegraphics[width=4in]{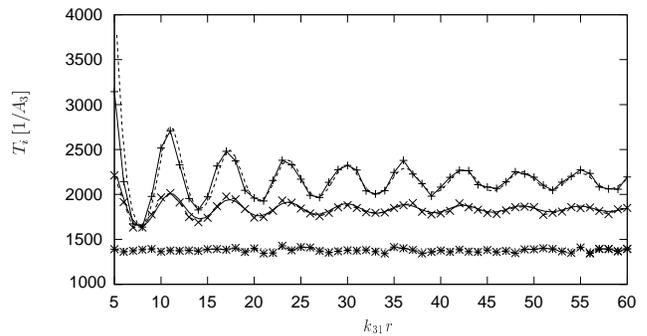}
\caption{Mean duration of fluorescence periods. Simulation: $T_2$~~ $+++$,~ 
$T_1$~~ $\times\times\times$,~ $T_0$~~ $\ast\ast\ast$. Theory: $T_2$~~ $- -
-$,~ $T_1$~~ $-\!\!\!\!-\!\!\!-\!\!\!-$,~ 
$T_0$~~ $....$ ~~~corrected for averaging window
($\Omega_3\!=\!0.5~A_3$, $\Omega_2\!=\!0.01~A_3$, zero
detuning).
\label{T012Verglneu160}} 
\end{figure}

To compare this with simulated data, obtained with a moving
averaging window of length $\Delta T_{\rm w}= 247~A_3^{-1}$, we have taken
$\Delta \tau=\frac{2}{3}\Delta T_{\rm w}$, as in the previous section, and have
plotted the results together with the simulated  data in
Fig. \ref{T012Verglneu160}. 
The agreement is very good.
Quite generally, for zero detuning the oscillations of $T_1$ and $T_2$
are {\em opposite} in phase to those of Re$\,C_3(r)$, as already noted at
the end of Section \ref{pij}. As in the case of the double jump rate, 
  $T_1$ and $T_2$ can become constant in $r$ for particular values 
of the detuning (different for $T_1$ and $T_2$ ),  and then change
to a  behavior {\em in phase} with Re$\,C_3(r)$.

The above approach of taking the  averaging window into account
works for the following reason. For a single atom with macroscopic
dark periods it is known that the emission of photons is describable,
to high accuracy, by an underlying two-step telegraph process. For two
independent atoms  with macroscopic dark periods the emissions are
therefore described by an underlying three-step telegraph process. For
two atoms interacting by a weak dipole-dipole interaction the actual
emission process of photons should therefore still have, at least
approximately, an underlying three-step telegraph process. What we
have done above is replacing the actual emission process by this
underlying three-step telegraph process and then incorporating the
averaging window by taking into account the influence of the
overlooked short periods  on the statistics.

\section{Discussion of results}\label{disc}
We have investigated cooperative effects in the fluorescence of two
dipole-dipole interacting atoms in a $V$ configuration. One of the
excited states of the $V$ configuration is assumed to be metastable,
i.e. with a weak transition to the ground state. When driven by two
lasers, a single such  configuration exhibits macroscopic dark periods and
periods of fixed intensity, like a two-step telegraph process. A
system of two such atoms exhibits three fluorescence  types,
i.e. dark periods and periods of single and double intensity, like a
three-step telegraph process. For large atomic distances, when the
dipole-dipole interaction is negligible, the total fluorescence
just consists  of the sum of the individual atomic contributions. We
have shown  that for
smaller atomic distances the fluorescence  modified by the 
dipole-dipole interaction which depends on the atomic distance $r$. In
particular we have, to our knowledge for the first time,
explicitly demonstrated cooperative effects in the rate of double
jumps from a period of double intensity to a
dark period or vice versa, both analytically and by simulations.

By means of an analytical theory we have obtained the r-dependent
transition rates , $p_{ij}$, between the three intensity
periods. These were then 
used to  calculate  the rate of double jumps and 
in the mean period durations $T_0,T_1,$ and $T_2$. When comparing with
the simulations  it turned
out that one had to take into account the averaging window used for
obtaining an intensity curve from the individual photon
emissions. With this the agreement between simulation and analytic
theory became excellent.

For zero laser detuning, for which the simulations were
performed, the  double
jump rates are {\em  in phase} with and  $T_1$ and $T_2$  {\em opposite}
in phase to Re$\,C_3(r)$. The theoretical expressions, however, allow
general detuning, $\Delta_2$, of the laser which drives the weak 
transition. It has been shown that for a particular $\Delta_2$, which
depends on the other 
parameters, the double jump rate becomes constant and, for larger
$\Delta_2$, varies opposite in phase to Re$\,C_3(r)$. A similar change of
characteristic behavior also occurs for $T_1$ and $T_2$, for
different values of $\Delta_2$ though. The amplitude of the oscillations with
the atomic distance remain in the same region of magnitude as for zero
detuning.  
As pointed out in Ref. \cite{BeHe5}, a dependence of the oscillations
on Re$\,C_3(r)$ is not unexpected since Re$\,C_3(r)$ affects the decay
rates of the excited Dicke states of the combined system. But an
intuitive argument why the above change of behavior occurs for
increased detuning  is at present not apparent.

\begin{figure}[bth]
\hspace{-2cm}\includegraphics[width=4in]{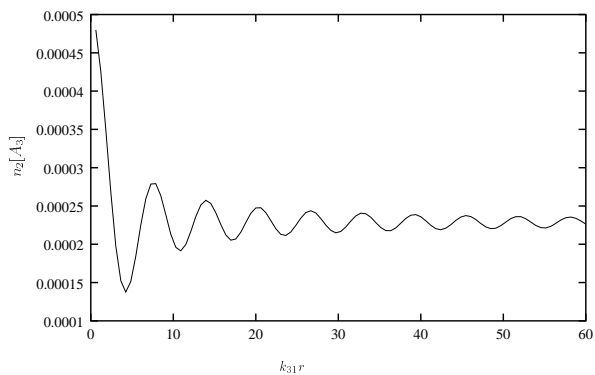}
\caption{The theoretical rate, $n_2$, of double intensity periods per unit time
   shows distance-dependent cooperative effects
  ($\Omega_3\!=\!0.5~A_3$, $\Omega_2\!=\!0.01~A_3$, zero detuning).
\label{n2}}
\end{figure}

We have pointed out in Section \ref{DJTheory} that there is another
statistical property of
the fluorescence which can serve as  an indicator of the influence of
the dipole-dipole 
interaction and which is probably not too difficult to determine
experimentally. This quantity is the rate with which fluorescence periods of
definite type occur, in 
particular the rate of periods with double intensity. Our
theoretical results show that this rate behaves similar to the
double jump rate, as regards the variation with the atomic
distance, and an example is shown in Fig. \ref{n2}. 
This quantity is
probably much easier to measure than the double jump rate or the
mean duration $T_2$.

Our theoretical approach can be carried
over to other level configurations and to more than two atoms. For
given parameters the evaluation should be not too difficult. If,
however, one is interested in closed algebraic expressions the effort
will increase considerably with the number of atoms. In particular, it
would be interesting to apply our approach to the situation of the
experiment of Ref. \cite{Sauter} with its different level 
configuration and its  three ions in the trap.

\begin{appendix}
\section{Dipole-dipole interaction in the Bloch equations} \label{A}
\begin{figure}[bth]
\hspace{-2cm}\includegraphics[width=4in]{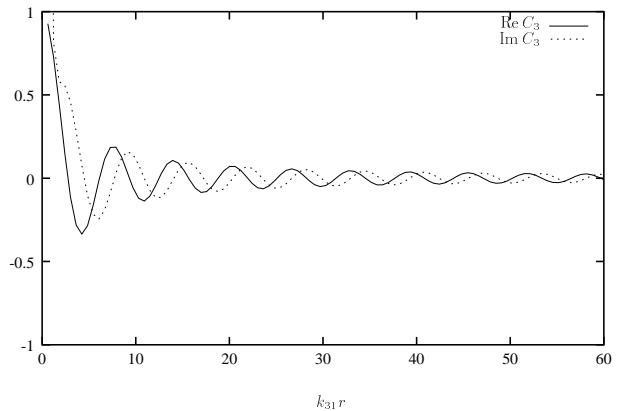}
\caption{The complex dipole-dipole coupling constant $C_3$ for the
  strong transition as a function of the atomic distance.
\label{C3}}
\end{figure}

The dipole-dipole interaction enters the Bloch equations through
$r$-dependent complex coupling constants (cf. Ref. \cite{BeHe5})
\begin{eqnarray} 
\label{Cj}
\lefteqn{
C_j= {3A_j \over 2} \, {\rm e}^{i k_{j1} r} \Bigg[
{1 \over {i}k_{j1} r} \left( 1 - \cos^2 \vartheta_j \right)}
\nonumber\\&&\hspace{5ex}
 + \left( {1 \over (k_{j1}r)^2} -{1 \over
     i(k_{j1}r)^3} \right)  
\left( 1 - 3 \cos^2 \vartheta_j \right) \Bigg].
\end{eqnarray}
Here $\vartheta_j$ is the angle between the transition dipole moment ${\bf
  D}_{1j}$ and the line connecting the atoms 
and $k_{j1}=2\pi/\lambda_{j1}$, where $\lambda_{j1}$ is the wavelength
of the j-1 transition for an atom. 
For $A_2 \approx 0$ one has $C_2 \approx 0$. Thus
one can neglect the dipole interaction when one atom is in state
$|2\rangle$. The dependence of $C_3$ on $r$ is maximal for
$\vartheta_3 = \pi/2$ 
and the corresponding $C_3$ is plotted in
Fig. \ref{C3}. 
For atomic distances greater than about three quarters
of a wave length of the strong transition, $|C_3|$ is 
less than $0.2\,A_3$, but for smaller distances Re$\,C_3$
approaches $A_3$ and Im$\,C_3$ diverges.

The reset operation $\cal R$ and $H_{\rm cond}$ are given 
by the same expressions as in Ref. \cite {BeHe5}, except for the
detuning.  One has 
\begin{equation} 
\label{calR}
{\cal R} (\rho)\!=\!\left(A_3\!+\!{\rm Re} \, C_3 \right) R_+ 
\rho R^\dagger_+\! +\! \left(A_3\! -\! {\rm Re} \, C_3 \right) R_-
\rho R_-^\dagger 
\end{equation}
where
\begin{eqnarray} \label{R+-}
R_+ &=& \left(S_{13}^- + S_{23}^- \right) / \sqrt{2}  \nonumber\\ 
&=& |g \rangle \langle s_{13}| \!+\! |s_{13} \rangle \langle e_3|
+ \big(|s_{12} \rangle \langle s_{23}|\! -\! |a_{12} \rangle \langle a_{23}| 
\big)/\sqrt{2},  \nonumber \\
R_- &=& \left(S_{13}^- - S_{23}^- \right) / \sqrt{2} \nonumber \\
&=&    |g \rangle \langle a_{13}| \!+\! |a_{13} \rangle \langle e_3| 
\!+\! \big(|s_{12} \rangle \langle a_{23}| \!+\! |a_{12} \rangle \langle
s_{23}| \big)/\sqrt{2} .
\end{eqnarray} 
The summands in Eq. (\ref{formhcond}) are given by
\begin{widetext}
\begin{eqnarray} 
\label{H0cond}
%\lefteqn{
H^0_{\rm cond}&=&  
 {\hbar   \over 2 i} \, \Big[  
A_3 \, \big( |s_{23} \rangle \langle s_{23}|  
+ |a_{23} \rangle \langle a_{23}| \big)
+   (A_3 + C_3) |s_{13} \rangle \langle s_{13}| 
+ ( A_3 - C_3) |a_{13} \rangle \langle a_{13}| 
+ 2 A_3 \, |e_3 \rangle \langle e_3|\Big] 
%}
\nonumber\\
&&\hspace{2ex}+{\hbar   \over 2 }\Big [\sqrt{2}  \Omega_3 \big( |g
\rangle \langle s_{13}|   
+ |s_{13} \rangle \langle e_3| \big) 
+ \Omega_3 \big( |s_{12} \rangle \langle s_{23}| 
- |a_{12} \rangle \langle a_{23}| \big) + {\rm H.c.}\Big ]\nonumber\\ 
&&\hspace{5ex}-\hbar \Delta_2\Big[2|e_2\rangle\langle e_2|
+|s_{12} \rangle \langle s_{12}| + 
|a_{12} \rangle \langle
a_{12}| + |s_{23} \rangle \langle s_{23}|  + |a_{23} \rangle \langle
a_{23}|\Big]
%} 
\end{eqnarray}
\begin{eqnarray}
H_{\rm cond}^1(\Omega_2)= {\hbar  \over 2 }\Big [
\sqrt{2}\Omega_2  \big( |g \rangle\langle s_{12}|+|s_{12} \rangle \langle e_2|
\big) 
+ \Omega_2 \big( |s_{13} \rangle \langle s_{23}| 
+ |a_{13} \rangle \langle a_{23}| \big) + {\rm H.c.}\Big ]  
\label{H0om}
\end{eqnarray} 
\end{widetext}
 From Eq. (\ref{H0cond}) one sees that Re$\,C_3$ changes 
the spontaneous decay rates and that Im$\,C_3$ leads to level
shifts.  Therefore, for small $r$, the decay rate of $|a_{13} \rangle$
approaches 0 in this case and the large level shifts cause a
decrease of fluorescence associated with the levels $|s_{13} \rangle$
and $|a_{13} \rangle$.

\section{calculation of $\rho(\lowercase{t}_0+\Delta \lowercase{t})$
  to first order in $\Omega_2$}

We write the Bloch equations of Eq. (\ref{Bloch}) in the form  
\begin{equation}\label{Bloch1}
\dot{\rho} = {\cal L}\rho   
\end{equation}
where the Liouvillean ${\cal L}\equiv {\cal
  L}(A_3,\Omega_3,\Delta_2,C_3,\Omega_2)$, a super-operator, can be
read off from Eqs. (\ref{Bloch}) and (\ref{calR}) - (\ref{H0om}). One
can  decompose ${\cal L}$ as 
\begin{equation}\label{Liouv}
 {\cal L} = {\cal L}_0+{\cal L}_{\Omega_2}   
\end{equation}
where ${\cal L}_0={\cal L}(A_3,\Omega_3,\Delta_2,C_3,0)$ and
${\cal L}_{\Omega_2}\rho =-{\rm i}[H_{\rm 
  cond}^1(\Omega_2),\rho ]/\hbar$. We note that $H_{\rm
  cond}^1(\Omega_2)$ is Hermitian and  that ${\cal L}_0$  can be
considered as a Liouvillean of Bloch equations. 
Choosing an
initial density matrix $\rho(t_0)$ lying 
in one of the subspaces in Eqs. (\ref{dark}) - (\ref{outer}) one obtains,
 to first order in $\Omega_2$,
\begin{eqnarray} 
\label{develop}
%\lefteqn{
\hspace{-5ex}\rho(t_0+\Delta t)&=&e^{{\cal L} \Delta t}\rho(t_0)\nonumber\\
\hspace{-5ex}&=&e^{{\cal L}_0 \Delta t}\rho(t_0)
%}
\nonumber\\&&\hspace{3ex}
+\int_{0}^{\Delta t}d\tau e^{{\cal L}_0
(\Delta t-\tau)}{\cal L}_{\Omega_2}e^{{\cal L}_0\tau}\rho(t_0),
\end{eqnarray}
just as with usual quantum mechanical perturbation theory in the interaction 
picture.  Now  we use the fact that ${\cal L}_0$, as a Liouvillean 
of Bloch equations, has an eigenvalue $0$ (corresponding to steady
states) and eigenvalues with negative real parts of  the order of
$\Omega_3$ and $A_3$. Therefore, if $\Delta t$ satisfies
Eq. (\ref{4.1}), the first term on the right-hand side of 
Eq. (\ref{develop}) gives one of the 
equilibrium states, $\rho^0$, of ${\cal L}_0$ given in Eqs. (\ref{ss0}) - 
(\ref{ss2}), to high accuracy, while the term $e^{{\cal L}_0
  \tau}\rho(t_0)$ under the integrand also rapidly approaches
$\rho^0$.    
After a change of integration variable  one therefore has to first
order in  $\Omega_2$  
\begin{equation}
\label{4.5}
\rho(t_0+\Delta t)=\rho^0 +
\int_{0}^{\Delta t}d\tau e^{{\cal L}_0\tau}
{\cal L}_{\Omega_2}\rho^0~.
\end{equation}
  It can be shown that 
${\cal L}_{\Omega_2}\rho^0$ has no components in the
zero-eigenvalue subspace   of ${\cal L}_0$ \cite{zero}. Therefore, the
integrand in 
Eq.~(\ref{4.5}) is rapidly 
damped, and since $\Delta t\gg \Omega_3^{-1},A_3^{-1}$, the upper
integration limit 
can  be extended to infinity. Hence we can write, to  first order in
$\Omega_2$, 
\begin{equation}
\label{4.6}
\rho(t_0+\Delta t)=\rho^0+
\int_{0}^{\infty}d\tau e^{{\cal L}_0\tau}
{\cal L}_{\Omega_2}\rho^0~.
\end{equation}
Thus,  if $\Delta t$ satisfies Eq. (\ref{4.1}) then, 
to  first order in $\Omega_2$, $\rho(t_0+\Delta t)$ is 
independent of $\Delta t$, and one has 
\begin{equation}
\label{4.7}
\rho(t_0+\Delta t)=\rho^0+(\epsilon-{\cal L}_0)^{-1}
{\cal L}_{\Omega_2}\rho^0
\end{equation}
to  first order in $\Omega_2$, where the limit $\epsilon\!\to +0$ is
understood.  Multiplying this by ${\cal L}-\epsilon$ gives
\begin{equation}
{\cal L}\rho(t_0+ \Delta t)= {\cal L}_{\Omega_2}(\epsilon-{\cal L}_0)^{-1}
{\cal L}_{\Omega_2}\rho^0={\cal O}(\Omega_2^2) 
\end{equation}
which is Eq. (\ref{decisive}). That the transition rates are
independent of the particular choice of $\Delta t$ follows from
Eqs. (\ref{4.7}) and (\ref{4.2}).

\end{appendix}


\begin{thebibliography}{99}
\bibitem{zero} If ${\cal L}_{\Omega_2}\rho^0$ had a
  zero-eigenvalue component this would 
 lead to a nonunique solution of  Eq. (\ref{mast3}), which is
not the case.
\end{thebibliography}
\end{document}